\documentclass[reprint,superscriptaddress,amsmath,amssymb,aps,prb]{revtex4-2}

\usepackage{graphicx} 
\usepackage{dcolumn} 
\usepackage{bm} 
\usepackage[version=4]{mhchem} 
\usepackage{siunitx} 
\usepackage{braket} 
\usepackage{multirow}

\begin{document}

\title{\textit{Ab initio} study of carrier mobility in \ce{Bi2O2Se}}

\author{Yubo Yuan}
\affiliation{School of Materials Science and Engineering, Zhejiang University, 
310027 Hangzhou, China.}
\affiliation{Department of Materials Science and Engineering, Westlake
  University, 310030 Hangzhou, China.}
\affiliation{Key Laboratory of 3D Micro/Nano Fabrication and Characterization of Zhejiang Province, 
School of Engineering, Westlake University, 310030 Hangzhou, China.}
\author{Ziye Zhu}
\affiliation{Department of Materials Science and Engineering, Westlake
  University, 310030 Hangzhou, China.}
\affiliation{Key Laboratory of 3D Micro/Nano Fabrication and Characterization of Zhejiang Province, 
School of Engineering, Westlake University, 310030 Hangzhou, China.}
\author{Jiaming Hu}
\affiliation{Department of Materials Science and Engineering, Westlake
  University, 310030 Hangzhou, China.}
\affiliation{Key Laboratory of 3D Micro/Nano Fabrication and Characterization of Zhejiang Province, 
School of Engineering, Westlake University, 310030 Hangzhou, China.}
\author{Wenbin Li}
\email{liwenbin@westlake.edu.cn}
\affiliation{Department of Materials Science and Engineering, Westlake
  University, 310030 Hangzhou, China.}
\affiliation{Key Laboratory of 3D Micro/Nano Fabrication and Characterization of Zhejiang Province, 
  School of Engineering, Westlake University, 310030 Hangzhou, China.}

\date{\today}

\begin{abstract}

\ce{Bi2O2Se} is an emerging high-performance layered semiconductor
with excellent stability. While experimental studies have explored
carrier transport across various doping levels for both $n$-type and
$p$-type conduction, a comprehensive theoretical understanding remains
incomplete. In this work, we present parameter-free first-principles
calculations of the electron and hole mobilities in \ce{Bi2O2Se}, based
on iterative solution of the Boltzmann transport equation that
includes electron-phonon scattering and ionized impurity scattering on an
equal footing. Intriguingly, we find that \ce{Bi2O2Se} exhibits high
electron mobilities in both the in-plane and out-of-plane directions,
whereas the hole mobilities are only significant in the in-plane
direction, displaying a unique three-dimensional (3D) electron
transport and two-dimensional (2D) hole transport behavior. At 300~K,
the calculated intrinsic electron and hole mobilities along the
in-plane direction are 447~$\mathrm{cm^2\,V^{-1}\,s^{-1}}$ and
29~$\mathrm{cm^2\,V^{-1}\,s^{-1}}$, respectively, which are primarily
affected by Fr\"ohlich electron-phonon interactions. 
Due to its large static dielectric permittivity, 
\ce{Bi2O2Se} exhibits an exceptionally high
low-temperature electron mobilities above
$1.0\times10^5~\mathrm{cm^2\,V^{-1}\,s^{-1}}$, and its electron mobilities
above 50~K is robust against ionized impurity scattering over a
wide range of impurity concentrations. By incorporating the Hall
effect into our analysis, we predict an in-plane electron Hall
mobility of 517~$\mathrm{cm^2\,V^{-1}\,s^{-1}}$ at 300~K, in excellent
agreement with experimental data. These results provide
valuable insights into the carrier transport mechanisms in
\ce{Bi2O2Se}, and offer predictive benchmarks for future theoretical
and experimental investigations.
\end{abstract}
\maketitle
\clearpage

\section{\label{sec:level1}introduction}

Layered semiconductors have attracted significant research attention
recently due to their promising applications in electronics and
optoelectronics devices~\cite{Liu2021_Nature,Kim2024_NatNano}.  Among them,
\ce{Bi2O2Se} stands out as a layered semiconductor with excellent air
stability and notable electronic and optical
properties~\cite{Li2021_AccMaterRes,Li2024_Infomat,Tan2025_NatRev}.  The bulk form of \ce{Bi2O2Se} is held together by weak interlayer electrostatic interactions and
can be cleaved into thin two-dimensional (2D)
layers~\cite{Chen2018_SciAdv}. Experimentally, thin layers of $n$-type \ce{Bi2O2Se}
typically exhibit high room-temperature electron mobility ($\mu>200~\mathrm{cm^2~V^{-1}~s^{-1}}$) when the layer thickness exceeds $6~\mathrm{nm}$, and the low-temperature electron mobility can reach $10^4~\mathrm{cm^2~V^{-1}~s^{-1}}$ or even higher~\cite{Wu2017_NatNano, Chen2018_SciAdv,Tan2022_NanoLet,Zhang2023_NatMater,Tan2023_Nature,Tang2025_NatMater}. 
On the other hand, intrinsic point defects can drive \ce{Bi2O2Se} into a
highly insulating state, with the Fermi level located below the charge
neutrality point, indicating $p$-type
behavior~\cite{Wang2023_NanoRes}. Notably, $p$-type \ce{Bi2O2Se}
transistors have been successfully fabricated via substitutional
doping with \ce{Zn^{2+}}, achieving a field-effect hole mobility up to
34 $\mathrm{cm^2~V^{-1}~s^{-1}}$ at 300 K~\cite{Wang2025_NatCommun}.

Despite these promising experimental results, a rigorous theoretical
understanding of carrier transport in \ce{Bi2O2Se}, particularly
regarding accurate mobility predictions, remains incomplete. 
The electron-phonon interaction plays a key role in determining the carrier transport properties of semiconductors. Previous DFT-based studies have 
employed approaches of various approximation levels—such as the Bardeen-Shockley deformation potential theory for bulk, thin-layer, and nanoribbon structures 
of \ce{Bi2O2Se}, and the semiclassical Boltzmann transport theory 
within the constant relaxation-time approximation—to investigate 
the phonon-limited electronic transport properties of 
\ce{Bi2O2Se}~\cite{Yu2018_APL,Wang2018_NJP,Wang2019_PCCP,Huang2019_APLMaterials,Huang2022_APL}. It should be noted that the deformation potential theory assumes acoustic phonons as the main source of carrier scattering, which is in general not valid in compound semiconductors with non-zero Born effective charges, as the contributions of optical phonons to carrier scattering are neglected~\cite{Ponce2020_review}. In particular, polar optical phonons in compound semiconductors can generate long-range electrical potentials that couple strongly to electron motion~\cite{Verdi2015_PRL}. Moreover, the carrier relaxation time in semiconductors is in general a function of both carrier energy and momentum, and a constant relaxation-time approximation is insufficient for accurate mobility predictions.
Besides electron-phonon interaction, electron-ionized impurity 
interaction also plays a critical role in determining the carrier transport properties of doped semiconductors\cite{Lu2022_PRM,Leveillee2023_PRB}. 
A prior study by some of the
authors~\cite{Zhu2022_JACS} used the fully \textit{ab initio} Boltzmann transport equation (BTE) within the self-energy relaxation time
approximation (SERTA) to calculate the phonon-limited electron 
mobilities of \ce{Bi2O2Se}, where the carrier lifetimes due to electron-phonon coupling were calculated fully from first principles, while a semi-empirical Brooks-Herring model was used to calculate momentum-dependent ionized-impurity scattering. The later was combined with the BTE to calculate the 
the ionized-impurity-limited electron mobility. Total electron mobilities, accounting for both electron-phonon scattering and electron-ionized impurity scattering, were then estimated via Matthiessen's rule. This study revealed that Fr\"ohlich
interaction has a strong impact on electron transport in \ce{Bi2O2Se}. Furthermore, the study revealed that the exceptionally high in-plane static dielectric constant ($\epsilon_0 >150$) of \ce{Bi2O2Se}~\cite{Xu2021, Zhu2022_JACS, Zhu2025_NanoLett, Zich2025_PRM} leads to effective screening of Coulomb scattering potentials from ionized impurities in the material, contributing to its ultrahigh low-temperature electron mobility and defect-tolerant electron mobility at room temperature. Recent experimental progress in fabricating $p$-type \ce{Bi2O2Se}~\cite{Wang2025_NatCommun} further motivated the study of hole transport in \ce{Bi2O2Se}. However, a first-principles investigation of hole mobility and transport mechanism in \ce{Bi2O2Se}
is still lacking. 

Moreover, recent years have seen notable progress in
predicting the carrier mobility of semiconductors with greater accuracy and
comprehensiveness.  For example, the inclusion of dynamical
quadrupole corrections were found to be important for accurate modeling of electron-phonon interactions in the carrier transport of certain semiconductors~\cite{Brunin2020_PRL,Jhalani20_PRL,Ponce2021_PRR,Ponce2023_PRB}.  With respect to the numerical solution of the BTE, although the use of Cauchy-distributed sampling at the Brillouin zone center for both
$\mathbf{k}$-points and $\mathbf{q}$-points in SERTA enables more
efficient calculations, SERTA can exhibit considerable errors when compared with more accurate iterative solution of the BTE for polar semiconductors~\cite{Claes2022_PRB}. In addition, a first-principles treatment of ionized impurity scattering t
accounts for the band structures, Kohn-Sham orbital overlaps, and
anisotropic dielectric constants have recently been developed and implemented in the EPW code~\cite{Leveillee2023_PRB}, opening opportunities for more accurate modeling of ionized-impurity scattering in semiconductors.

Leveraging these methodological advances, in this work we present a detailed first-principles investigation of electron and hole mobilities in bulk \ce{Bi2O2Se}, utilizing a state-of-the-art iterative solution of the BTE that incorporates quadrupole
corrections. Experimentally, mobility measurements for \ce{Bi2O2Se} were usually carried out on bulk or multilayer samples with a thickness above $6~\mathrm{nm}$, corresponding to $\sim$10 layers~\cite{Wu2017_NatNano,Chen2018_SciAdv,Tan2022_NanoLet}. For such a layer number, the intrinsic mobility of layered materials typically have already reached the corresponding bulk limit~\cite{Li2019_NanoLett}. Therefore, our study of the transport properties of bulk \ce{Bi2O2Se} are still highly pertinent to the understanding of the experimental results. We further compare the calculated electron mobility to those obtained from approximate solutions of the BTE (such as SERTA), for a temperature range between 10 K to 400 K. At low
temperatures, the total electron mobilities of \ce{Bi2O2Se} including both electron-phonon scattering and ionized-impurity scattering
were predicted for electron concentrations from
$10^{14}~\mathrm{cm^{-3}}$ to $10^{17}~\mathrm{cm^{-3}}$. We find that
\ce{Bi2O2Se} exhibits intrinsic three-dimensional (3D) electron and 2D hole transport,
with polar optical phonons dominating the scattering processes for both electron and hole transport at room temperature. We further predict in-plane electron Hall mobility of
\ce{Bi2O2Se}, which agrees well with a previous experimental result.


\section{\label{sec:level2}methods}
\subsection{Computational details}
For the calculations of the basic electron and phonon properties of \ce{Bi2O2Se}, we employed the density functional theory (DFT) and density functional perturbation 
theory (DFPT), as implemented in the Quantum ESPRESSO 
package~\cite{Baroni2001_RevModPhy,Giannozzi2017_JPCM}. 
The generalized gradient approximation (GGA) of Perdew, Burke, and 
Ernzerhof (PBE) was used for the exchange-correlation functional~\cite{Perdew1996_PRL}. 
Optimized norm-conserving Vanderbilt pseudopotentials were obtained 
from the PseudoDojo library~\cite{Hamann2013_PRB,Setten2018_CPC}. 
A plane-wave cutoff energy of 80~Ry was used for all calculations. 
The Brillouin zone was sampled using an $8 \times 8 \times 8$ 
$\mathbf{k}$-point grid, and a $4 \times 4 \times 4$ phonon 
$\mathbf{q}$-point grid was employed for DFPT calculations.

Calculations of electron-phonon coupling and carrier mobility were performed using the EPW package~
\cite{Giustino2007_PRB,Lee2023_npj,Ponce2016_CPC,Noffsinger2010_CPC},
and the maximally localized Wannier function method was applied using
the Wannier90 software~\cite{Pizzi2020_JPCM}. Carrier mobilities for both
electrons and holes were computed by solving the iterative
BTE~\cite{Ponce2018_PRB,Macheda2018_PRB}.  The BTE method is
computationally expensive, as it requires a uniform fine grid size
across the Brillouin zone. The dynamical quadrupole tensor of
\ce{Bi2O2Se} was evaluated using DFPT as implemented in the ABINIT
code~\cite{Royo2019_PRX, Gonze2020_CPC, Gonze1997_PRB_1,
  Gonze1997_PRB_2, Romero2020_JCP, Gonze2016_CPC}. To compute Hall
mobility~\cite{Macheda2018_PRB, Ponce2021_PRR}, a magnetic field of
$10^{-10}$~T was applied on \ce{Bi2O2Se}.  Adaptive broadening was
employed to approximate Dirac delta functions~\cite{Ponce2021_PRR}.
The carrier concentration of \ce{Bi2O2Se} was set to
$10^{14}~\mathrm{cm^{-3}}$ for calculations of intrinsic phonon-limited 
mobility. To capture the effect of ionized impurity 
scattering~\cite{Leveillee2023_PRB}, the doping concentration was further
increased to $10^{17}~\mathrm{cm^{-3}}$. 
We employ the monopole approximation to describe the potential of an impurity of charge $Ze$, which was set to $Z = 1$ throughout the calculations.

For the convergence of intrinsic electron mobility,
in the temperature range of 200--400~K, an energy window of 0.3~eV 
was applied and a fine Brillouin zone grid of up to $200^3$ was used. 
To achieve the convergence of electron mobilities at lower temperatures 
(below 200~K), a smaller energy window of 0.1~eV suffices (Fig.~S1~\cite{SupplyInformation}). In this case, 
the densest fine grid reaches $420^3$ to converge electron mobility at 10~K, and the result was linearly extrapolated to the limit of an infinitely dense Brillouin zone grid. 
The small slopes of the linear fitting lines indicate good convergence (Fig.~S2). 
Electron mobilities as the electron concentration ranged from $10^{14}~\mathrm{cm^{-3}}$ 
to $10^{17}~\mathrm{cm^{-3}}$ were also linearly extrapolated with 
respect to the size of the grid (Fig.~S3). The convergence of the fine grid size for 
the hole mobility calculations in \ce{Bi2O2Se} was achieved with a 
grid size of up to $90^3$, using an energy window of 0.3~eV (Fig.~S4). 

\begin{figure*}[th!]
\includegraphics[scale=1]{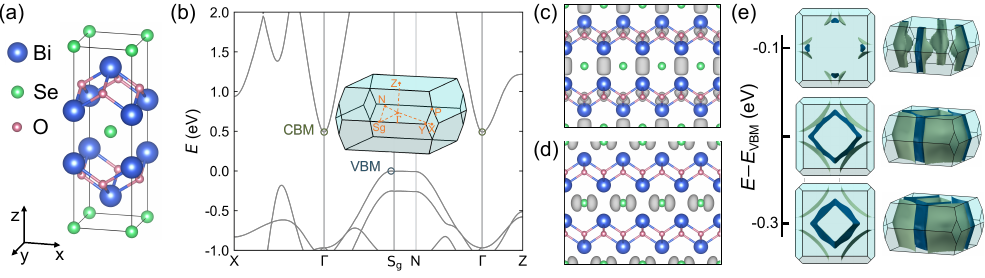}
\caption{\label{fig:One} 
(a) Atomistic structure of \ce{Bi2O2Se} in a tetragonal conventional unit cell. 
(b) Electronic band structure of \ce{Bi2O2Se}, with the conduction 
band minimum (CBM) and valence band maximum (VBM) labeled. 
Inset shows the reciprocal space path employed in the band structure.
(c) Isosurfaces of the electron wave function at the CBM (gray). 
(d) Isosurfaces of the hole wave function at the VBM (gray). 
(e) Evolution of the Fermi surfaces at energies of 0.1, 0.2, and 0.3 eV 
below the VBM, shown from bird's-eye views and side views of the 
Brillouin zones. The front- and back-side colors of Fermi surfaces 
are blue and green, respectively.}
\end{figure*}

\subsection{Theory of carrier mobility calculations}
The low-field electron mobility is given by the following expression 
(a similar form applies for hole mobility)~\cite{Ponce2020_review}:
\begin{equation}
    \label{eq:ph-mob}
        \mu _ { \alpha \beta } = 
        \frac { -1 } { \Omega_\mathrm{uc} n _ { \mathrm {c} } } 
        \sum _ {n} \int \frac { \mathrm {d}^{3} \mathbf{k} } 
        { \Omega _ { \mathrm {BZ} } } v _ { n \mathbf {k} \alpha } 
        \partial _ { E _ { \beta } } f _ { n \mathbf { k } },
\end{equation}
where the Greek indices $\alpha$ and $\beta$ run over the three Cartesian directions. 
The symbol $n_{\mathrm{c}}$ refers 
to the carrier concentration. The quantity $v_{n \mathbf{k} \alpha}$ represents the 
band velocity along the $\alpha$ direction for the Kohn-Sham state with band index $n$ and wavevector $\mathbf{k}$. The volumes of 
the unit cell and the first Brillouin zone are represented by $\Omega_\mathrm{uc}$ and 
$\Omega_{\mathrm{BZ}}$, respectively.
$\partial_{E_\beta} f_{n \mathbf{k}}$ corresponds 
to the linear response of the electronic occupation of the state $n \mathbf{k}$ to the applied electric field $E$. Within the framework of the linearized BTE, 
$\partial_{E_\beta} f_{n \mathbf{k}}$ is derived as follows~\cite{Ponce2020_review}:
\begin{equation}
    \label{eq:eleOcc}
    \begin{aligned}
        \partial _ { E _ { \beta } } f _ { n \mathbf {k} } &= 
        e v _ { n \mathbf {k} \beta } \frac { \partial f _ 
        { n \mathbf {k} } ^ { 0 } } { \partial \epsilon _ 
        { n \mathbf{k} } } \tau _ { n \mathbf{k} } \\
        &+ \frac { 2 \pi \tau _ { n \mathbf{k} } } { \hbar } 
        \sum _ { m \nu } \int \frac { \mathrm{d} ^ {3} \mathbf{q} } 
        { \Omega _ { \mathrm {BZ} } } | g _ { m n \nu } 
        ( \mathbf{k} , \mathbf{q} ) | ^ { 2 } \\
        &\times [ ( n _ { \mathbf{q} \nu } + 1 - f _ { n \mathbf{k} } 
        ^ {0} ) \delta ( \epsilon _ { n \mathbf{k} } - 
        \epsilon _ { m \mathbf{k} + \mathbf{q} } + 
        \hbar \omega _ {\mathbf{q} \nu} ) \\
        &+ ( n _ { \mathbf{q} \nu } + f _ {n \mathbf{k}} ^ {0} ) 
        \delta ( \epsilon _ { n \mathbf{k} } - 
        \epsilon _ { m \mathbf{k} + \mathbf{q} } - 
        \hbar \omega _ { \mathbf{q} \nu} ) ] \partial _ { E _ { \beta } } 
        f _ { m \mathbf{k} + \mathbf{q} } ,
    \end{aligned} 
\end{equation}
where $e$ is the elementary charge, $f^0_{n \mathbf{k}}$ represents the Fermi-Dirac 
occupation of the state $n \mathbf{k}$ in the absence of an external electric field. 
The electron-phonon matrix elements $g_{mn\nu}(\mathbf{k}, \mathbf{q})$ describe 
the amplitude for an electron to scatter from the initial state $n \mathbf{k}$ to the 
final state $m \mathbf{k}+\mathbf{q}$ through the emission or absorption of a phonon with 
frequency $\omega_{\mathbf{q} \nu}$, and $\epsilon_{n \mathbf{k}}$ and 
$\epsilon_{m \mathbf{k} + \mathbf{q}}$ denote the Kohn-Sham eigenenergies of the respective states. 
The Bose-Einstein distribution is given by $n_{\mathbf{q} \nu}$, 
and $\tau_{n\mathbf{k}}$ denotes the scattering lifetime. 

To reduce the computational cost of solving the BTE, the SERTA simplifies the BTE by 
neglecting the second term in Eq.(\ref{eq:eleOcc}). This approximation assumes that 
scattering out of a state $n \mathbf{k}$ is dominant and neglects scattering back into it. 
The momentum relaxation time approximation (MRTA) partially accounts for these neglected 
back-scattering processes, providing a more accurate estimate of transport properties 
than SERTA, though still an approximation to the full iterative BTE solution. 
~\cite{Ponce2020_review}.
The inverse of the partial scattering lifetime, $\tau ^ {-1} _ {n\mathbf{k} \rightarrow m\mathbf{k} + \mathbf{q}}$, 
corresponds to the partial scattering rate and is given by the sum of 
the electron–phonon (ph) rate and electron–impurity (imp) scattering rate:
\begin{equation}
    \label{eq:scatterRate}
          \tau ^ {-1} _ {n\mathbf{k} \rightarrow m\mathbf{k} + \mathbf{q}} = \tau ^ \mathrm{-1, \mathrm{ph} }_ 
          {n\mathbf{k} \rightarrow m\mathbf{k} + \mathbf{q}} + \tau ^ \mathrm{-1, \mathrm{imp} } _ { n\mathbf{k} \rightarrow m\mathbf{k} + \mathbf{q} }.
\end{equation}
where the partial electron–phonon scattering rate is defined as:
\begin{equation}
    \label{eq:ph-tau}
    \begin{aligned}
        \tau _ { n \mathbf{k} \rightarrow m\mathbf{k} + \mathbf{q} } ^ {-1, \mathrm{ph} } = 
        &\frac { 2 \pi } { \hbar } \sum _ { \nu } | 
        g _ { m n \nu } ( \mathbf {k} , \mathbf {q} ) | ^ { 2 } 
        [ ( n _ { \mathbf { q } \nu } + 1 - 
        f _ { m \mathbf { k + q } } ^ { 0 } ) \\
        &\times \delta ( \epsilon _ { n \mathbf{k} } - 
        \epsilon _ { m \mathbf { k } + \mathbf { q } } - 
        \hbar \omega _ { \mathbf { q } \nu  } ) \\
        &+ ( n _ { \mathbf { q } \nu }  
        + f _ { m \mathbf { k + q } } ^ { 0 } ) 
        \delta ( \epsilon _ { n \mathbf { k } } - 
        \epsilon _ { m \mathbf { k } + \mathbf { q } } + 
        \hbar \omega _ { \mathbf { q } \nu } ) ] .
    \end{aligned} 
\end{equation}

For electron-ionized impurity scattering, the Kohn-Luttinger ensemble averaging method can be employed to compute 
the partial scattering rate due to randomly distributed 
ionized impurities~\cite{Leveillee2023_PRB}:
\begin{equation}
    \label{eq:ii-mob}
    \begin{aligned}
        \tau_{n\mathbf{k} \rightarrow m\mathbf{k} + \mathbf{q} } ^ {-1, \mathrm{imp} } = 
        & n _ { \mathrm{imp} } \frac { 2 \pi } { \hbar } 
        \left[ \frac { e ^ { 2 } } { 4 \pi \varepsilon _ { 0 } } 
        \frac { 4 \pi Z } { \Omega _ { \mathrm{uc} } } \right] ^ { 2 } \\
        &\times \sum _ { \mathbf{G} \not = \mathbf{-q} } \frac { | 
        \Braket{\psi _ { m \mathbf{k} + \mathbf{q}}|e ^ { i (\mathbf{q}+\mathbf{G}) 
        \cdot \mathbf{r} }| \psi _ { n \mathbf{k} } }_\mathrm{uc} | ^ { 2 } } 
        { | ( \mathbf{q} + \mathbf{G} ) \cdot \mathbf{\epsilon}_{ 0 } 
        \cdot ( \mathbf{q} + \mathbf{G} ) | ^ { 2 } } \\
        &\times \delta ( \epsilon_{ n \mathbf{k} } - 
        \epsilon_{ m \mathbf{k} + \mathbf{q} } ),
    \end{aligned} 
\end{equation}
where $n_\mathrm{imp}$ and $Z$ denote the concentration and charge of 
ionized impurities. The $\varepsilon_0$ is the vacuum permittivity.
The tensor $\epsilon_0$ is the low-frequency dielectric constant, 
accounting for both electronic and ionic polarizability. 
The calculated dielectric constants of \ce{Bi2O2Se} are 156 in-plane and 83 out-of-plane, 
as obtained using the Quantum ESPRESSO package. These values are slightly lower than 
those reported in our previous study~\cite{Zhu2022_JACS}, where the dielectric constants 
were computed directly from analytical equations. However, this small difference arises 
from methodological variations and does not affect the validity or accuracy of the 
current transport calculations.

The Hall mobility is calculated under the influence of a small external magnetic field $\mathbf{B}$, which exerts a Lorentz force on the electrons, thereby affecting their mobility. The Hall mobility is expressed as:
\begin{equation}
    \label{eq:ph-Hall-mob}
          \mu _ { \alpha \beta \gamma} ^ { \mathrm { Hall} } = 
          \frac { -1 } { \Omega_\mathrm{uc} n _ { \mathrm {c} } } 
          \sum _ {n} \int \frac { \mathrm {d}^{3} \mathbf{k} } 
          { \Omega _ { \mathrm {BZ} } } v _ { n \mathbf {k} \alpha } 
          [\partial _ { E_{\beta} } f _ { n \mathbf { k } }(B_{\gamma}) 
          - \partial _ { E_{\beta} } f _ { n \mathbf { k } }] ,
\end{equation}
where $\gamma$ denotes the Cartesian components of the magnetic field. To obtain the Hall mobility, the BTE is solved as:
\begin{equation}
    \label{eq:Hall-mob}
    \begin{aligned}
        \left[ 1 - \frac { e } { \hbar } \tau _ { n \mathbf{k} } 
        ( \mathbf{v} _ { n \mathbf{k} } \times \mathbf{B} ) 
        \cdot \nabla _ { \mathbf { \mathbf{k} } } \right] 
        \partial _ { E _ { \beta } } f _ { n \mathbf {k} } (B_{\gamma}) = 
        e v _ { n \mathbf {k} \beta } \frac { \partial f _ 
        { n \mathbf {k} } ^ { 0 } } { \partial \epsilon _ 
        { n \mathbf{k} } } \tau _ { n \mathbf{k} } \\
        + \frac { 2 \pi \tau _ { n \mathbf{k} } } { \hbar } 
        \sum _ { m \nu } \int \frac { \mathrm{d} ^ {3} \mathbf{q} } 
        { \Omega _ { \mathrm {BZ} } } | g _ { m n \nu } 
        ( \mathbf{k} , \mathbf{q} ) | ^ { 2 } \\
        \times [ ( n _ { \mathbf{q} \nu } + 1 - f _ { n \mathbf{k} } 
        ^ {0} ) \delta ( \epsilon _ { n \mathbf{k} } - 
        \epsilon _ { m \mathbf{k} + \mathbf{q} } + 
        \hbar \omega _ {\mathbf{q} \nu} ) \\
        + ( n _ { \mathbf{q} \nu } + f _ {n \mathbf{k}} ^ {0} ) 
        \delta ( \epsilon _ { n \mathbf{k} } - 
        \epsilon _ { m \mathbf{k} + \mathbf{q} } - 
        \hbar \omega _ { \mathbf{q} \nu} ) ] \partial _ { E _ { \beta } } 
        f _ { m \mathbf{k} + \mathbf{q} }(B_{\gamma}) .
    \end{aligned} 
\end{equation}

The Hall mobility is directly related to the drift mobility through the Hall factor, 
which is defined as:
\begin{equation}
    \label{eq:Hall-factor}
          r^\mathrm{Hall}_{\alpha \beta \gamma}
          = \frac { \mu _ { \alpha \beta \gamma} ^ { \mathrm { Hall} } } 
          {\mu _ { \alpha \beta }}. 
\end{equation}
The Hall factor depends on both the material properties and temperature~\cite{Claes2025_NatRevPhy}.

\section{\label{sec:level3}results and discussion}

\ce{Bi2O2Se} has a layered tetragonal structure in the conventional cell (Fig.~1(a)).
In \ce{Bi2O2Se}, the conduction band minimum (CBM) is located at the $\Gamma$ point, 
as shown in Fig.~1(b). The second conduction band valley lies 400 meV above the CBM. 
At 400~K (the highest temperature considered in this study), the average thermal energy  
($\frac{3}{2}k_\mathrm{B}T$) corresponds to 52~meV, indicating that the available thermal 
energy is insufficient to promote electrons into the second valley. As a result, there 
are no accessible final states 
for electron intervalley scattering. Therefore, intravalley scattering dominates electron 
transport under low electric fields. The electronic bands near the CBM are primarily 
composed of Bi 6\textit{p} orbitals, whereas those near the valence band maximum (VBM) 
mainly originate from O 2\textit{p} and Se 4\textit{p} orbitals (Fig.~S5). Thus, 
Wannier functions including all of these \textit{p} orbitals can fully 
capture the 
relevant bands (Fig.~S6). Bi \textit{p} orbital connectivity across layers reveals a 
three-dimensional character in the electron conducting states (Fig.~1(c)). The states 
at the VBM are primarily derived from Se \textit{p} orbitals towards the in-plane direction 
(Fig.~1(d)). The ellipsoidal electron pocket near the CBM is localized at the center of the 
Brillouin zone (Fig.~S7)~\cite{Lv2019_PRB}. However, as the eigenenergies near the VBM decrease, 
the hole pockets in the Brillouin zone initially appear near the four edges, 
and then interconnect and evolve into split, warped square shapes, as shown in Fig.~1(e). 
The shapes of the calculated Fermi surfaces near the VBM are consistent with those 
previously observed by angle-resolved photoemission spectroscopy~\cite{Liang2019_AdvMater}.

We employ first-principles calculations to predict the intrinsic phonon-limited electron 
mobility Fig.~2 presents the electron mobilities from 10~K to 400~K.
The calculated BTE electron mobilities along both the out-of-plane and in-plane 
directions remain high across the entire temperature range, indicating 3D electron 
transport. We further compare the exact BTE solution with approximation methods. 
The in-plane mobilities obtained using the MRTA and SERTA were tested for convergence with 
respect to the $\mathbf{k}$- and $\mathbf{q}$-point grid size, which were randomly distributed 
around the $\Gamma$ point with a Cauchy distribution (Fig.~S8). When the number of fine 
$\mathbf{k}$- and $\mathbf{q}$- point grid size reaches 
$\sim$35000, the in-plane mobilities 
calculated using MRTA show good agreement with the BTE results over the temperature 
range of 50--400~K. However, below 50~K, the electron mobilities predicted by MRTA tend to be lower than those obtained from the full BTE. In contrast, the SERTA consistently underestimates the in-plane electron mobilities across the entire temperature range.

\begin{figure}[tb!]
\includegraphics[scale=1]{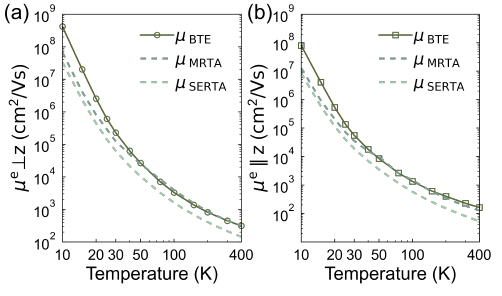}
\caption{\label{fig:four}
(a) Temperature dependence of the in-plane intrinsic phonon-limited electron mobilities.
In-plane mobilities are compared with momentum relaxation-time approximation (MRTA) 
and self-energy relaxation-time approximation (SERTA).
(b) Temperature dependence of the out-of-plane intrinsic phonon-limited electron mobilities.
Out-of-plane mobilities are compared with momentum relaxation-time approximation (MRTA) 
and self-energy relaxation-time approximation (SERTA).}
\end{figure}

\begin{figure}[tb!]
\includegraphics[scale=1]{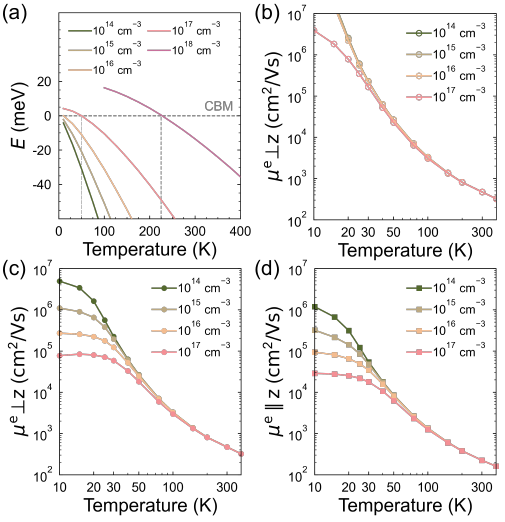}
\caption{\label{fig:iis_mob} 
(a) Temperature dependence of the Fermi level for ionized impurity 
concentrations ranging from $10^{14}~\mathrm{cm^{-3}}$ to $10^{18}~\mathrm{cm^{-3}}$, 
with the conduction band minimum set to zero.
(b) Temperature dependence of in-plane intrinsic phonon-limited electron mobilities.
(c) Temperature dependence of in-plane total electron mobilities, 
including both phonon and ionized impurity scattering.
(d) Temperature dependence of out-of-plane total electron mobilities, 
including both phonon and ionized impurity scattering.
In (b), (c), and (d), ionized impurity concentrations range from 
$10^{14}~\mathrm{cm^{-3}}$ to $10^{17}~\mathrm{cm^{-3}}$.}
\end{figure}

\textit{N}-type conductivity in \ce{Bi2O2Se} can be induced by 
intrinsic defects or ionized impurities ~\cite{Ding2022_Matter,Zhan2015_JECR}.
As the electron concentration increases, \ce{Bi2O2Se} gradually becomes a 
degenerate semiconductor, with the Fermi level shifting into the conduction 
band~\cite{Xu2021_SciChina}. 
Fig. 3(a) shows that the temperature at which the \ce{Bi2O2Se} becomes a
degenerate semiconductor decreases from 225 K to 50 K as the doping 
concentration is reduced from $10^{18}~\mathrm{cm^{-3}}$ to 
$10^{17}~\mathrm{cm^{-3}}$.
\ce{Bi2O2Se} remains semiconducting above 10~K at lower electron concentrations.
In the metallic regime for electron concentrations equal to or greater 
than $10^{18}~\mathrm{cm^{-3}}$, the plasmon energy exceeds the vibrational 
energy of the Fröhlich longitudinal optical (LO) phonon mode, more 
sophisticated modeling of free-carrier screening is necessary to 
predict phonon-limited electron mobility accurately~\cite{Verdi2017_NatCom,Zhu2022_JACS}. 
Consequently, we limit our investigation to the low-temperature electron mobilities over a  
doping range from $10^{14}~\mathrm{cm^{-3}}$ to $10^{17}~\mathrm{cm^{-3}}$.
In Figure~3(b), at a carrier concentration of $10^{17}~\mathrm{cm^{-3}}$, the phonon-limited 
electron mobility exhibits a noticeable decline, which can be attributed to 
the decrease of electron average mean free path due to more 
efficient electron-phonon scattering rates. 

Nevertheless, at low temperatures, ionized impurity scattering becomes a dominant factor 
influencing electron mobility, in addition to electron-phonon scattering~\cite{Zhu2022_JACS}. To 
further investigate the total electron mobilities, we consider both phonon and ionized 
impurity scattering mechanisms, as shown in Fig.~3(c) and Fig.~3(d). The high total electron mobilities are indicative of weak ionized impurity scattering, 
due to the strong dielectric screening of the Coulomb potential from ionized impurities~\cite{Zhu2022_JACS}. 
It is important that plateaus in total mobilities are observed at low temperatures 
(around 10--30~K) as the electron concentration increases, 
consistent with trends reported in experimental studies~\cite{Wu2017_NatNano}. 
At 10~K and an electron concentration of $10^{17}~\mathrm{cm^{-3}}$, 
the in-plane and out-of-plane total mobilities reach $7.7\times10^4$ and 
$2.9\times10^4~\mathrm{cm^2~V^{-1}~s^{-1}}$, respectively. The total electron mobilities 
of \ce{Bi2O2Se} remain nearly unchanged above 50~K regardless of the ionized impurity
concentration, primarily due to the dominance of electron-phonon scattering 
over ionized impurity scattering. This behavior contrasts with that of silicon 
and other conventional semiconductors with typical low static dielectric constants, 
where ionized impurities can significantly reduce the total electron mobility~\cite{Leveillee2023_PRB}. At room temperature, our full BTE calculations yield electron mobilities of 447~$\mathrm{cm^2~V^{-1}~s^{-1}}$ along the in-plane direction  and 227~$\mathrm{cm^2~V^{-1}~s^{-1}}$ along the out-of-plane direction.

\begin{figure}[tb!]
\includegraphics[scale=1]{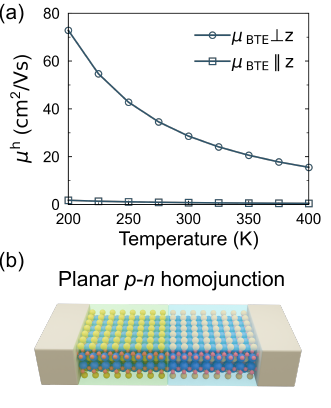}
\caption{\label{fig:hole}
(a) Temperature dependence of the in-plane and out-of-plane hole 
mobilities when the carrier concentration is $10^{14}~\mathrm{cm^{-3}}$. 
(b) Schematic diagram illustrating a planar \textit{p-n} junction of \ce{Bi2O2Se}.}
\end{figure}

\begin{figure}[htb]
\includegraphics[scale=0.8]{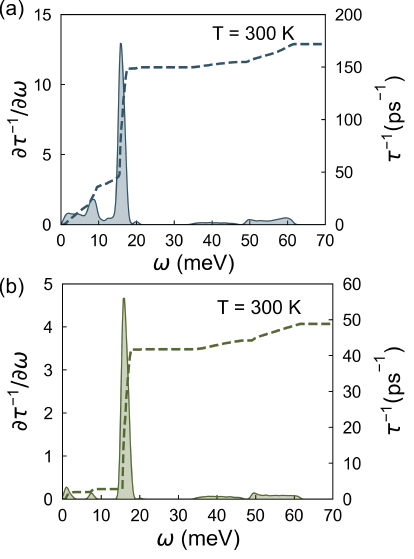}
\caption{\label{fig:spectralDecomp}
(a) Spectral decomposition of the angularly averaged electron scattering rates as 
a function of phonon energy at 300~K, calculated using a fine $200^3$ $\mathbf{k}$- 
and $\mathbf{q}$-point grid size. 
(b) Spectral decomposition of the angularly averaged hole scattering rates 
as a function of phonon energy at 300~K, calculated using a fine $90^3$ $\mathbf{k}$- 
and $\mathbf{q}$-point grid size. 
In (a) and (b), dashed lines indicate the cumulative integrals of the scattering rates 
(axes on the right).}
\end{figure}

Next, we turn to the hole mobility. As shown in Figure~4(a), 
the in-plane hole mobility at room temperature is predicted to be 29~$\mathrm{cm^2~V^{-1}~s^{-1}}$, 
while the out-of-plane hole mobility is nearly zero.
Although spin-orbit coupling (SOC) can significantly impact hole mobilities in some 
semiconductors even with relatively small coupling strengths~\cite{Ponce2021_PRR}, 
it has a negligible effect on the band dispersion of \ce{Bi2O2Se} near the valence band 
edges and has a negligible effect on the hole transport properties of \ce{Bi2O2Se} (Fig.~S9). 
Specifically, comparison with the BTE solutions including SOC shows that the in-plane 
electron mobilities computed without SOC exhibits an average error of only 5.6\% over 
the range of 200--400~K. Considering the excellent in-plane electron and 
hole mobilities, we propose a planar \textit{p-n} homojunction of \ce{Bi2O2Se}, 
as illustrated in Fig.~4(b).

Fig.~5(a) presents the spectral decomposition of angularly averaged 
electron-phonon scattering rates as a function of phonon frequency at 300~K, 
evaluated for electrons located 39~meV 
($\frac{3}{2}k_\mathrm{B}T$) above the CBM. 
The total electron scattering rate is found to be 49~$\mathrm{ps^{-1}}$. 
High-energy optical phonon branches (above 30~meV) contribute weakly to 
electron scattering across a broad frequency range. In contrast, three 
peaks are identified below 20~meV, corresponding to the longitudinal 
acoustic (LA), transverse optical (TO), and LO phonon modes, listed in 
order of increasing frequency. The LO phonon mode contributes approximately 
80\% of the total scattering, highlighting the dominant role of the Fröhlich 
interaction in electron-phonon scattering. By comparison, the LA and TO modes 
contribute only 4.0\% and 1.7\%, respectively. To assess the effect of the 
dynamical quadrupole correction, we recompute the electron scattering rates 
without including it. In this case, the total scattering rate becomes 
48~$\mathrm{ps^{-1}}$, consistent with our prior estimate of 50~$\mathrm{ps^{-1}}$
obtained using Cauchy sampling, also neglecting quadrupole 
corrections~\cite{Zhu2022_JACS}. The contributions from the LA and TO modes are  
reduced to 1.7\% and 0.8\%, respectively, as shown in Fig.~S10. 
Previous studies~\cite{Ponce2021_PRR, Brunin2020_PRL, Wang2024_APL} 
have demonstrated that in certain polar materials, quadrupole corrections 
predominantly influence acoustic rather than optical phonons. In the case 
of \ce{Bi2O2Se}, the small contribution of acoustic modes to the total 
scattering explains why electron mobilities remain largely unaffected by the quadrupole 
corrections. Nevertheless, the relative importance of these low-frequency 
acoustic and optical modes is expected to increase at low temperatures, 
where the population of LO phonons is significantly reduced.
It should be noted that some previous studies reported the change 
from polar optical phonon (POP) scattering to piezoelectric scattering 
at low temperatures in few-layer flakes, which can be associated 
with the polar nature of the 2D ferroelectric material~\cite{Yang2021_AM,Yip2024_nanotech}.
This behavior is distinctly different from what we calculate 
for the bulk 3D phase in the paraelectric (PE) state.
It possesses inversion symmetry in its PE state. 
As a result, the piezoelectric tensor is identically zero, 
and piezoelectric scattering is absent.

For the electron-phonon scattering mechanism of the hole carriers, we present the spectral decomposition of angularly averaged hole scattering rates 
by phonon frequencies at 300 K in Fig.~5(b), for holes at an energy of 39 meV 
below the VBM. The LO phonon mode shows strong coupling with holes, 
accounting for 60\% of the total scattering rates. At lower frequencies, 
the spectral decomposition exhibits broad peaks, making it difficult to 
identify specific phonon modes. To analyze these modes in more detail, 
we plot the mode resolved electron-phonon matrix elements of \ce{Bi2O2Se} 
with the initial state at the VBM in Fig.~S11. The LA phonon mode 
interacts with holes across a wide frequency range (0--12~$\mathrm{meV}$), 
indicating that intravalley and intervalley scattering mediated by LA 
phonons contributes to the hole scattering. The total hole scattering 
rate is 172~$\mathrm{ps^{-1}}$. 
The ultrasmall out-of-plane hole mobility is primarily attributed to the large 
out-of-plane hole effective mass ($13.24m_0$). In contrast, the in-plane hole 
effective mass ($m^*$) is significantly smaller, with a value of $0.96m_0$ along 
the $\mathrm{S_g}$--$\Gamma$ path and $0.19m_0$ along the direction perpendicular 
to the $\mathrm{S_g}$--$\Gamma$ path.
The harmonic average of the in-plane hole effective mass ($\overline{m^*}$) is $0.31m_0$. 
Using the simple Drude formula ($\mu = e/(\overline{m^*} \cdot \tau^{-1})$), 
the calculated in-plane hole mobility is 33~$\mathrm{cm^2~V^{-1}~s^{-1}}$, 
which agrees well with our BTE results.

Finally, Hall mobility is generally measured over drift mobility 
for comparison with experimental data. Hence, Hall effect should be 
taken into account accordingly~\cite{Ponce2021_PRR}. To ensure 
numerical accuracy, the Hall factor was linearly extrapolated 
with respect to the Brillouin zone grid size. The Hall factor is 
anisotropic, with the in-plane component being slightly higher 
than the out-of-plane component (Fig.~S12). By combining the previously 
calculated electron drift mobilities with the Hall factor, the Hall 
mobilities can be obtained, as shown in Fig.~6. 
At 300~K, the calculated in-plane electron Hall mobility is 
517~$\mathrm{cm^2~V^{-1}~s^{-1}}$. This value is in good agreement 
with the experimental Hall mobility of 440~$\mathrm{cm^2~V^{-1}~s^{-1}}$
measured in a sample with a comparably low residual carrier 
concentration~\cite{Wu2019_NanoLet}.
We do not compare with other experimental data, as most of the data are 
from a high-doping regime beyond the scope of our current model~\cite{Lin2025_small}.


\begin{figure}[tb!]
\includegraphics[scale=1]{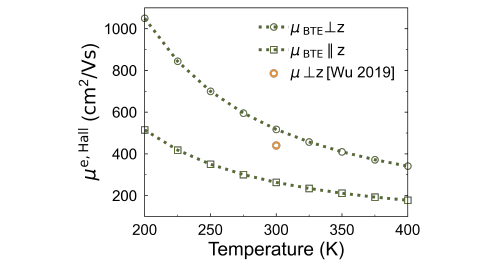}
\caption{\label{fig:Hall_mob} 
Temperature dependence of electron Hall mobilities 
along the in-plane direction and along 
the out-of-plane direction from 200 K to 400 K. The experimental 
in-plane electron Hall mobility is at an electron concentration 
of $3\times10^{17}~\mathrm{cm^{-3}}$ obtained from Ref.~\cite{Wu2019_NanoLet}.}
\end{figure}

\section{\label{sec:level4}conclusions}
In summary, we have carried out a comprehensive first-principles
investigation of electron and hole transport in \ce{Bi2O2Se},
employing an accurate iterative solution to the BTE with quadrupole
corrections. Despite its layered structure, \ce{Bi2O2Se} exhibits
excellent electron transport in both the in-plane and out-of-plane
directions. Approximate methods such as SERTA and MRTA serve as
efficient alternatives to the full BTE solution. However, they
underestimate the in-plane electron mobilities, while MRTA results are
consistent with the BTE calculations in the temperature range of
50--400~K.  At low temperatures, we have predicted the total electron
mobilities in $n$-type \ce{Bi2O2Se} limited by both electron-phonon 
scattering and ionized impurity scattering for carrier concentrations ranging
from $10^{14}$ to $10^{17}~\mathrm{cm^{-3}}$. At higher temperatures
(200--400~K), electron-phonon scattering becomes the dominant mechanism. At 300~K, the intrinsic electron mobilities are predicted to be 447 and
227~$\mathrm{cm^2~V^{-1}~s^{-1}}$ for the in-plane and out-of-plane
directions, respectively. The in-plane hole mobility is
29~$\mathrm{cm^2~V^{-1}~s^{-1}}$, while the out-of-plane hole mobility
is nearly zero, owing to the large out-of-plane effective mass of
holes. The efficient 3D electron and 2D hole transport suggest the
feasibility of a planar \textit{p-n} homojunction for future high-performance
transistors. Moreover, the computed in-plane electron Hall mobility is
517~$\mathrm{cm^2~V^{-1}~s^{-1}}$, aligned with experimental data.  This
work provides significant insights into the carrier transport and
scattering mechanisms in the high-performance layered semiconductor
\ce{Bi2O2Se}, offering predictive benchmarks for the future
development of \ce{Bi2O2Se}-based electronic and optoelectronic
applications.

\section*{Appendix A: Dynamical quadrupole tensors}
The computed dynamical quadrupole tensors for \ce{Bi2O2Se} are listed in Table~1. 
The Se site, being inversion symmetric, has a vanishing quadrupole tensor. 
In contrast, the Bi and O sites lack inversion symmetry and thus exhibit generally 
nonzero components. Specifically, the Bi1 and O1 atoms transform into Bi2 and O2 
aligned along the $z$ direction under reflection across the $xy$-plane (see Fig.~S14). 
Under this reflection, each occurrence of a $z$ index in a tensor component introduces 
a sign change. For a given quadrupole tensor component $Q^{\kappa}_{\alpha\beta\gamma}$, 
where $\kappa$ indexes atoms in the primitive cell of \ce{Bi2O2Se} and the rest of Greek indexes ($\alpha$, $\beta$, and $\gamma$) indicate Cartesian directions, the relation 
between Bi1 (or O1) and Bi2 (or O2) is:
\[Q^{\mathrm{Bi2/O2}}_{\alpha\beta\gamma} = (-1)^n\,Q^{\mathrm{Bi1/O1}}_{\alpha\beta\gamma} ,\]
where $n$ is the number of $z$ indices among $\alpha, \beta, \gamma$. 
Thus, quadrupole tensor components with an odd $n$ change sign. 
The total sum of the quadrupole tensor components over all atoms in \ce{Bi2O2Se} vanishes, 
consistent with the absence of a third-rank piezoelectric tensor, which is subject to the 
same symmetry constraints. This is a consequence of the $I4/mmm$ space group and the 
presence of inversion symmetry in the crystal structure.
Dynamical quadrupole corrections are included in all subsequent carrier mobility 
calculations unless stated otherwise.

\begin{table}[htb!]
    \centering
    \caption{\label{tab:quadrupole}
    Quadrupole tensor ($Q^{\kappa}_{\alpha\beta\gamma}$) of \ce{Bi2O2Se} (in e$\cdot$Bohr).}
    \setlength{\tabcolsep}{1mm}
    \begin{tabular}{cccccccc}
        \hline
        \hline
        \multirow{2}*{$\kappa$} & \multirow{2}*{$\alpha$~} & \multicolumn{6}{c}{${\beta\gamma}$} \\
        \cline{3-8} 
         &  & $xx$ & $yy$ & $zz$ & $yz,zy$ & $xz,zx$ & $xy,yx$  \\
        \hline
             & $x$ &         0.00 &         0.00 &        0.00 &        0.00 & 8.06&        0.00\\
        Bi1  & $y$ &         0.00 &         0.00 &        0.00 & 8.06 &         0.00&        0.00\\
             & $z$ &11.34   &11.34 & 0.94 &        0.00 &         0.00&        0.00\\
        \hline
             & $x$ &          0.00 &          0.00 &         0.00 &         0.00 & -8.06&         0.00 \\
        Bi2  & $y$ &          0.00 &          0.00 &         0.00 &-8.06&         0.00 &         0.00 \\
             & $z$ &-11.34 &-11.34 &-0.94 &         0.00 &         0.00 &         0.00 \\
        \hline
             & $x$ &         0.00 &          0.00 &         0.00 &          0.00 & 13.59 &         0.00 \\
        O1   & $y$ &         0.00 &          0.00 &         0.00 & -13.59 &         0.00 &         0.00 \\
             & $z$ &   10.68 & -10.68 &         0.00 &          0.00 &         0.00 &         0.00 \\
        \hline
             & $x$ &         0.00 &          0.00 &         0.00 &         0.00 & -13.59 &         0.00 \\
        O2   & $y$ &         0.00 &          0.00 &         0.00 & 13.59 &          0.00 &         0.00 \\
             & $z$ & -10.68 & 10.68 &         0.00 &         0.00 &          0.00 &         0.00 \\
        \hline
             & $x$ &         0.00 &         0.00 &         0.00 &         0.00 &         0.00 &         0.00 \\
        Se   & $y$ &         0.00 &         0.00 &         0.00 &         0.00 &         0.00 &         0.00 \\
             & $z$ &         0.00 &         0.00 &         0.00 &         0.00 &         0.00 &         0.00 \\
        \hline
        \hline
    \end{tabular}
\end{table}

\section*{Acknowledgments}
This work is supported by National Natural Science Foundation of China (Grant No. 62374136) and the National Key R\&D Program
of China (2024YFA1409600). The authors thank the
High-Performance Computing Center of Westlake University for technical
assistance.

\textit{Data availability.} The data that support the findings of 
this article are openly available~\cite{database}.

\bibliography{EPW.bib}

\end{document}